\documentclass[%
 aip,
 amsmath,amssymb,
 reprint,%
]{revtex4-1}

\usepackage{graphicx}
\usepackage{dcolumn}
\usepackage{bm}
\usepackage{svg}

\usepackage[utf8]{inputenc}
\usepackage[T1]{fontenc}
\usepackage{mathptmx}
\usepackage{etoolbox}
\usepackage[version=4]{mhchem}
\usepackage{braket}

\usepackage{siunitx} 
\usepackage{float} 
\usepackage[export]{adjustbox} 
\usepackage{mhchem}
\usepackage{hyperref}
\usepackage{miller}
\usepackage[caption=false]{subfig}
\usepackage{multirow}

\usepackage{xcolor}

\newcommand{\ttwo}{T$_2$}

\makeatletter
\def\@email#1#2{%
 \endgroup
 \patchcmd{\titleblock@produce}
  {\frontmatter@RRAPformat}
  {\frontmatter@RRAPformat{\produce@RRAP{*#1\href{mailto:#2}{#2}}}\frontmatter@RRAPformat}
  {}{}
}%
\makeatother
\begin{document}

\preprint{AIP/123-QED}

\title[Extended T$_2$ Times of Shallow Implanted NV in Chemically Mechanically Polished Diamond]{Extended T$_2$ Times of Shallow Implanted NV in Chemically Mechanically Polished Diamond}
\author{S. Tyler}
\affiliation{Department of Physics, University of Warwick, Coventry CV4 7AL, UK}
\author{J. Newland}
\affiliation{Element Six, Global Innovation Centre, Fermi Avenue, Harwell OX11 0QR, UK}
\author{P. Hepworth}%
\affiliation{Department of Physics, University of Warwick, Coventry CV4 7AL, UK}
\author{A. Wijesekara}
\affiliation{Department of Physics, University of Warwick, Coventry CV4 7AL, UK}
\author{I. R. Gullick}
\affiliation{Department of Physics, University of Warwick, Coventry CV4 7AL, UK}
\author{M. L. Markham}
\affiliation{Element Six, Global Innovation Centre, Fermi Avenue, Harwell OX11 0QR, UK}
\author{M. E. Newton}
\affiliation{Department of Physics, University of Warwick, Coventry CV4 7AL, UK}
\author{B.~L.~Green} 
\email[]{b.green@warwick.ac.uk}
\affiliation{Department of Physics, University of Warwick, Coventry CV4 7AL, UK}

\date{\today}

\begin{abstract}
Mechanical polishing of diamond is known to be detrimental to the spin lifetime and strain environment of near-surface defects. By utilising a chemical mechanical polishing (CMP) process, we demonstrate that we can achieve $^{13}\mathrm{C}$-limited spin lifetimes of shallow implanted ($\leq \SI{34}{\nano\meter}$) NV centres in an industrially scalable process. We compare spin lifetimes (T$_2$) of three diamonds processed with CMP with one processed by inductively-coupled plasma reactive ion etching (ICP-RIE), and observe an increased median T$_2$ of \SI{355}{\micro\second} in the CMP-processed samples for $^{15}$NV centres implanted and annealed under identical conditions. 
\end{abstract}

\maketitle

\section{Introduction}
Diamond has emerged as a leading candidate for quantum applications due to the prevalence of multiple extrinsic defects with combinations of exceptional photonic and spin properties. \cite{doherty2013,desantis2021,zhang2023,bradac2019} Of these, the negatively-charged nitrogen-vacancy (NV) center is the best studied and remains at the forefront of multiple quantum technology applications due to its high-fidelity optical spin initialization and readout. \cite{hopper2018} The applications of NV include  magnetic and electric field sensing at macro and nanoscales, as well as quantum information storage and processing. \cite{taylor2008,dolde2011,Neumann2010,kucsko2013,rondin2014}

In many of these applications the limiting property is the ground state electron spin lifetime, T$_2$, motivating many works dedicated to extending T$_2$ via improvements in material \cite{braunbeck,kim2014},  measurement protocols \cite{pham2012,morishita2019}, or a combination of the two. T$_2$ lifetimes are limited by interactions between the ground state electron spin and its environment. In high quality, low strain natural isotopic abundance material, the bulk \ttwo{} lifetime is limited by the \SI{1.1}{\percent} \ce{^{13}C} nuclear spin bath to approximately \SI{700}{\micro\second}. \cite{stephen2019,zhao2012} Imperfections in the material such as local impurities, extended defects, and other paramagnetic defects can quickly reduce this lifetime. \cite{zheng2022} 

Shallow NV centers are required for many high-sensitivity or nanoscale sensing applications, as the magnetic dipolar coupling between the sensor and the analyte $\propto 1/r^3$. \cite{belthangady2013} For functional near-surface sensors, the challenge is in maximizing the coupling between the sensor and the analyte whilst minimizing parasitic interactions (both magnetic and the local chemical potential, which affects the defects' charge state) with the diamond surface itself. Previous studies report the coherence time of shallow single NV's to be of the order $\leq$\SI{100}{\micro\second},\cite{t2times1,t2times2,sangtawesin} and this varies significantly with NV depth. A major environmental factor in reduced spin lifetimes for shallow implanted NV centres is sub-surface damage \cite{sangtawesin} which manifests as small cracks, dangling bonds, and trapped charges which can cause spin decoherence. The production of a ``perfect'' damage- and defect-free diamond surface is therefore a vital step in realizing the full potential of  NV-based platforms for quantum technologies. 

Traditionally, diamond surfaces are prepared by mechanical polishing, with high quality surface finishes typically being realized through the use of a process known as scaife polishing for e.g., optical and electronic applications. This process employs a cast iron wheel with embedded diamond grit and uses a mixture of catalytic sp$^3$-sp$^2$ transformation and diamond-on-diamond cutting to remove material. \cite{hird2004,pastewka2011} Significant anisotropy is observed for different polishing directions on different surfaces. \cite{luo2021} This hard polishing treatment inherently introduces sub-surface damage in the form of micro-fractures due to brittle contact between the sample and the polishing medium, making it unsuitable for the preparation of sensing-suitable surfaces. 

Multiple alternative approaches to achieving a low surface roughness, low damage surface have been studied in the literature, with the present incumbent being inductively-coupled plasma reactive ion etching (ICP-RIE). This process is capable of achieving RMS roughnesses of $\SI{\leq.2}{\nano\meter}$ over micron-scale areas whilst maximizing NV \ttwo{} lifetimes. \cite{sangtawesin} Care is typically taken to ensure that the etching process is not finalized using a chlorine-based process as there is some evidence that it is deleterious for near-surface NV properties. \cite{ruf2019,tao2014} Other investigated processes include UV-enhanced quartz polishing, \cite{watanabe2013} and chemical mechanical polishing (CMP). \cite{thomas2014-2} Initial investigations into the latter on single crystal diamond observed a minimal improvement in spin lifetimes. \cite{braunbeck} 

In this paper, we report the properties of near-surface implanted NV centers, and compare spin lifetimes for samples prepared using deep ICP-RIE etching, and a combination of ICP-RIE and CMP. We find that the mean spin lifetime for implanted NV is improved in the CMP-prepared samples for all implantation depths, revealing CMP to be a scalable route for producing state-of-the-art diamond surfaces.

\section{Method}
\begin{figure}[h]
\includegraphics[width=\columnwidth]{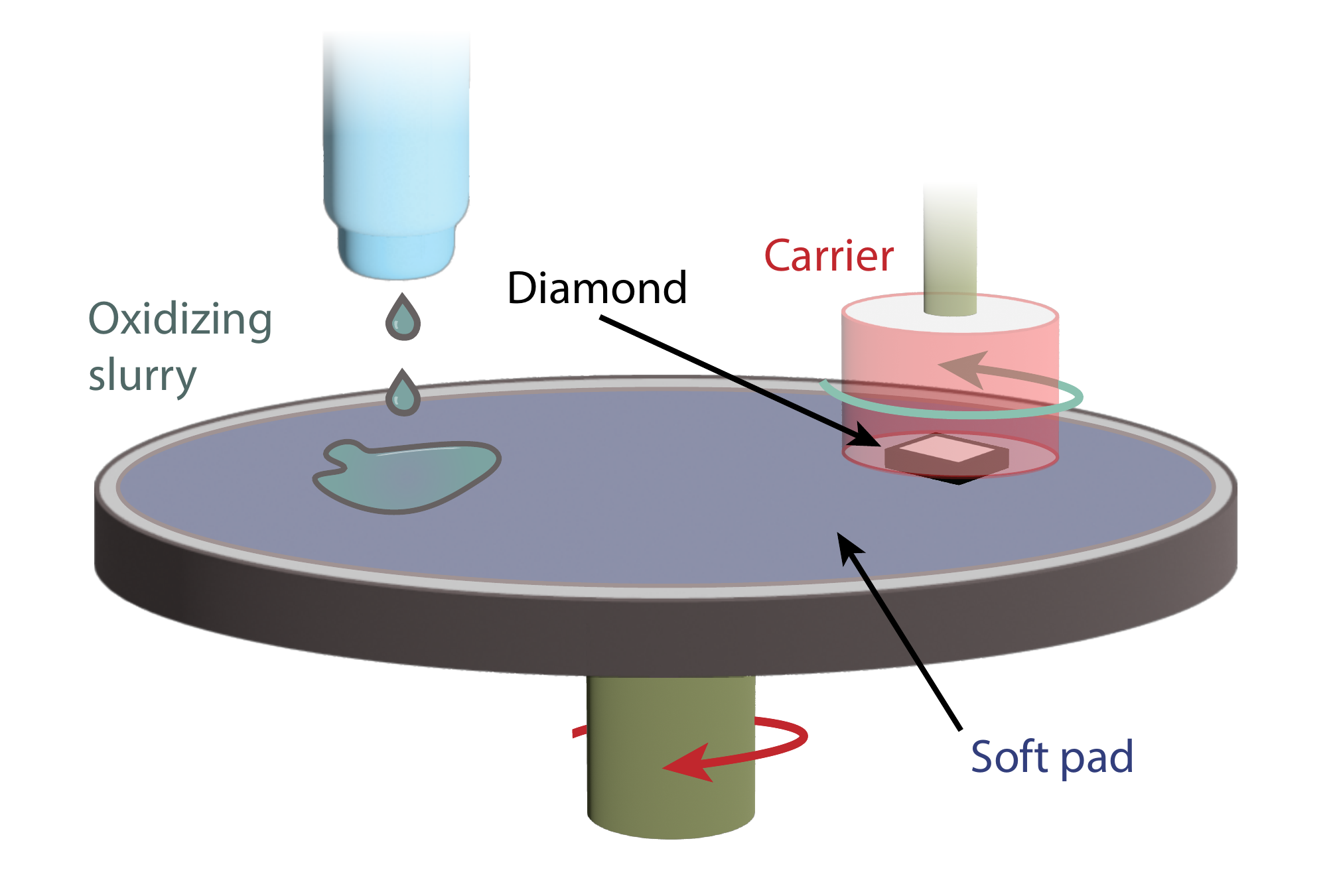}
\caption{\label{fig:cmp}In a CMP process, the abrasives contained within the oxidizing slurry and the pad itself are chosen to be softer than the surface of the diamond so no direct polishing can occur. Following oxidation, the softer material is subsequently removed by particulates embedded in the slurry and pad.}
\end{figure}

The CMP process is employed extensively in the electronics industry for silicon and other materials. \cite{zantye2004} It relies on an oxidizing slurry to transform the surface of the material being polished to a softer form, which can then be removed by mechanical abrasion (Fig.~\ref{fig:cmp}). The abrasion particles are chosen to be softer than the original material, so that no fracture or cutting occurs in the absence of the oxidizer. For diamond, the oxidizing solution of \ce{KMnO4} is chosen to promote the transformation of surface layers of sp$^3$-bonded carbon to sp$^2$-bonded carbon. Mechanical abrasion by \ce{Al2O3} and \ce{MnO2} abrasives then removes the sp$^2$ layer, resulting in an inherently subsurface-damage-free surface. 

In this work, five electronic-grade samples from Element Six were studied. Three samples were first etched using a pure \ce{O2} etch to remove approximately \SI{10}{\micro\meter} of material, and were then polished using CMP for \SI{50}{}-\SI{100}{\hour} to remove approximately \SI{1}{}-\SI{10}{\micro\meter} of material. A fourth sample was etched using a cyclical \ce{ArCl2}-\ce{O2} ICP-RIE etch based on the recipe described in \cite{ruf2019} with a cycle time of 5 minutes per stage and a cooldown period of 5 minutes between stages. A total of \SI{20}{\micro\meter} of material was removed from the scaife-finished surface: in both the CMP-processed and ICP-processed samples, the removed material should be sufficient to remove subsurface damage from the mechanical polishing.\cite{Hicks2019a} The final sample was analyzed in its as-received (scaife-finished) state for surface finish comparisons. 

Each of the CMP-processed samples was then implanted using $^{15}$N ions at a sample tilt of \SI{7}{\degree} and at energies of \SI{5}{keV}, \SI{10}{keV}, or \SI{20}{keV} to a dose of \SI{1e9}{\per\centi\meter\squared}: these energies are expected to result in mean implantation depths of \SI{9}{\nano\meter}, \SI{18}{\nano\meter}, and \SI{34}{\nano\meter} respectively, based on Crystal-TRIM \cite{posselt1994} calculations (see Appendix \ref{sec:sampleprep}). The predicted straggles are approximately \SI{3}{\nano\meter}, \SI{5}{\nano\meter}, and \SI{7}{\nano\meter} respectively. The ICP-RIE-processed sample was subsequently implanted with \SI{20}{keV} \ce{^15N} ions, under the rationale that, of the depths measured, this depth is expected to maximize the measured spin lifetimes. After implantation, the samples were buried in sacrificial diamond grit and annealed in a dry nitrogen atmosphere at \SI{400}{\celsius}, \SI{800}{\celsius}, and \SI{1200}{\celsius} for \SI{4}{\hour}, \SI{2}{\hour}, and \SI{2}{\hour} respectively. The samples were cleaned using piranha acid and subsequently the surface was terminated in an oxygen-containing atmosphere: full details of the sample preparation can be found in Appendix \ref{sec:sampleprep}. 

A home-built \SI{532}{nm} confocal microscope was used to image and investigate individual implanted NV centres: the nitrogen isotope and spin lifetime (T$_2$) were measured using Ramsey and Hahn echo-decay sequences, respectively. By taking the Fourier transform of each Ramsey measurement, the nitrogen isotope can be identified through the number of hyperfine transitions. The natural isotopic abundance of \ce{^15N} is \SI{0.37}{\percent}: \cite{junk1958} we therefore assume that measured \ce{^15NV} centers are direct products of implantation, incorporating the implanted ion; \ce{^14NV} centers are either native or produced by migration of implant-induced vacancies to pre-existing nitrogen impurities. \cite{vandam2019} Full measurement protocols and instrument details are given in Appendix~\ref{sec:confocal}. 


\section{Results \& Discussion}

\subsection{Surface finish}
In addition to low subsurface damage, many quantum technology applications also require low surface roughness for e.g., further processing, or integration with other hybrid technologies. A significant amount of work has been performed on the effect of CMP on nano- and ultranano-crystalline diamond, \cite{thomas2014,klemencic2017,werrell2017} but relatively little has been reported on single crystal finishes. \cite{liao2021}

\begin{figure}[h]
\includegraphics[scale=0.75]{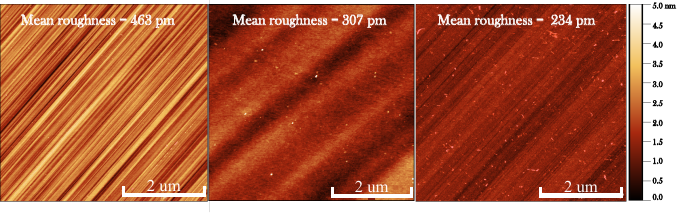}
\caption{\label{fig:afm} AFM images of (a) a scaife polished diamond sample, (b) a diamond treated with ICP-RIE and (c) a CMP treated diamond, alongside their respective mean surface roughness ($
S_a$).}
\end{figure}

The surface finish result of each surface preparation was analysed using a series of \SI{5}{\micro\meter} $\times$ \SI{5}{\micro\meter} atomic force microscopy (AFM) images from various regions on the surface of each sample (Fig.~\ref{fig:afm}).

The scaife-finish sample displays fine, parallel lines characteristic of well-mechanically-polished single crystal diamond surfaces, with the parallel grooves following the soft polishing direction for the \hkl<100> surface. \cite{hird2004} Samples prepared by the cyclical ICP-RIE etch and those prepared by the combined ICP+CMP processing have a reduced surface roughness compared to the scaife-finished sample. The cyclical ICP-RIE etched sample shows a smoothing of the scaife-finished lines characteristic of the \ce{ArCl2} process.\cite{ruf2019} The pure \ce{O2} etch utilized before CMP typically doesn't have the same smoothing effect, and hence the scaife polishing lines are still visible in the ICP+CMP-processed sample. Nevertheless, the combined process has reduced the magnitude of the grooves and resulted in the lowest RMS roughness surface, demonstrating that CMP is capable of producing low roughness surfaces. 

\begin{figure}[H]
\includegraphics[scale=0.5]{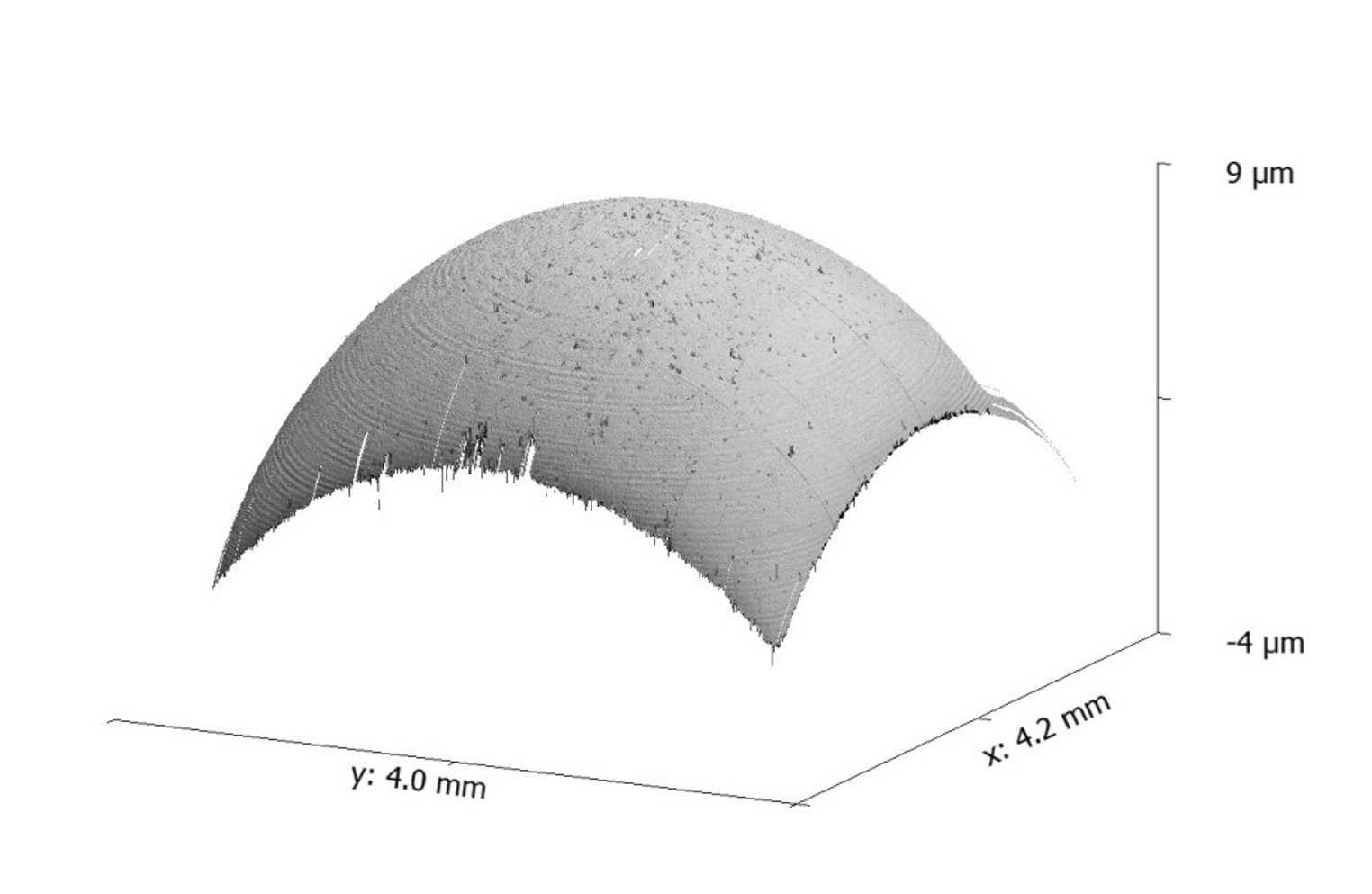}
\caption{\label{fig:rounding} Rounding of the sample's surface edges due to the use of a soft polishing pad during the CMP process. Increased accessibility of the sample edges to the oxidizing and abrasive slurry combined with a soft pad which distorts under the applied load results in preferential removal of material at the edges and corners.}
\end{figure}

One notable effect of CMP on the surface topography is rounding of the sample (Fig.~\ref{fig:rounding}), caused by the softness of the polishing pad which molds around the sample and results in increased material removal at the edges of the sample, which are most accessible to the oxidizing slurry. This intrinsic rounding of the sample surface will be prohibitive in some potential technological applications of \ce{NV} and further work is required to mitigate or eliminate this effect to maximize the potential of CMP processing of diamond. 

\subsection{Spin lifetimes}
Spin lifetime measurements on the ICP-RIE and ICP+CMP-processed samples were undertaken with a static field of \SIrange{170}{310}{G} aligned to one of the four NV orientations.\cite{doherty2013} Many of the shallow implanted NV centres demonstrated blinking during measurements caused by charge switching between \ce{NV^-} and \ce{NV^0}. This effect is most pronounced for shallow NV where carrier interactions with the surface can result in defect charge instability. \cite{fu2010,hauf2011} Unexpectedly, this effect was particularly prevalent with the sample implanted with an energy of \SI{10}{keV}, despite not containing the NV centres in the closest proximity to the surface. In order to rectify this, the sample was re-terminated with oxygen multiple times using a variety of piranha clean, \ce{O2} plasma, and \ce{O2} furnace termination. However, no improvement in the stability of the NV centres in this sample was observed irrespective of the termination method: it is possible that this is a result of a property of the sample itself, rather than the surface. Therefore, limited data was collected from the sample implanted at \SI{10}{keV}, and we do not report on it further. 

Spin lifetimes were measured by automating the confocal microscope to iterate over \ce{NV} centers in a confocal image using Qudi control software. \cite{qudi2017} For each NV, a quick ODMR spectrum was collected to confirm defect orientation relative to the static magnetic field, a Hanbury-Brown-Twiss (HBT) measurement confirmed each center was optically isolated, and optical power saturation and spin lifetime measurements were then performed. In each sample, many more NV centres containing $^{14}$N were sampled than those containing $^{15}$N. It is likely that this is a result of the vacancies produced during implantation being trapped by the existing nitrogen concentration (\SI{\approx1}{ppb}, versus a peak concentration of implanted \ce{^{15}N} \SI{\approx5}{ppb} [Fig.~\ref{fig:trim}]) combined with a higher probability of implanted centers displaying charge instability. \cite{rodgers} Data from both \ce{^{15}NV} and \ce{^{14}NV} are included here as a comparison between implanted and native nitrogen. We note that the background \ce{NV} concentration is essentially zero away from the implantation depth and therefore all centers are assumed to be products of implantation either directly via ion introduction, or through vacancy capture. 

\begin{figure}[H]
    \includegraphics[width=\columnwidth]{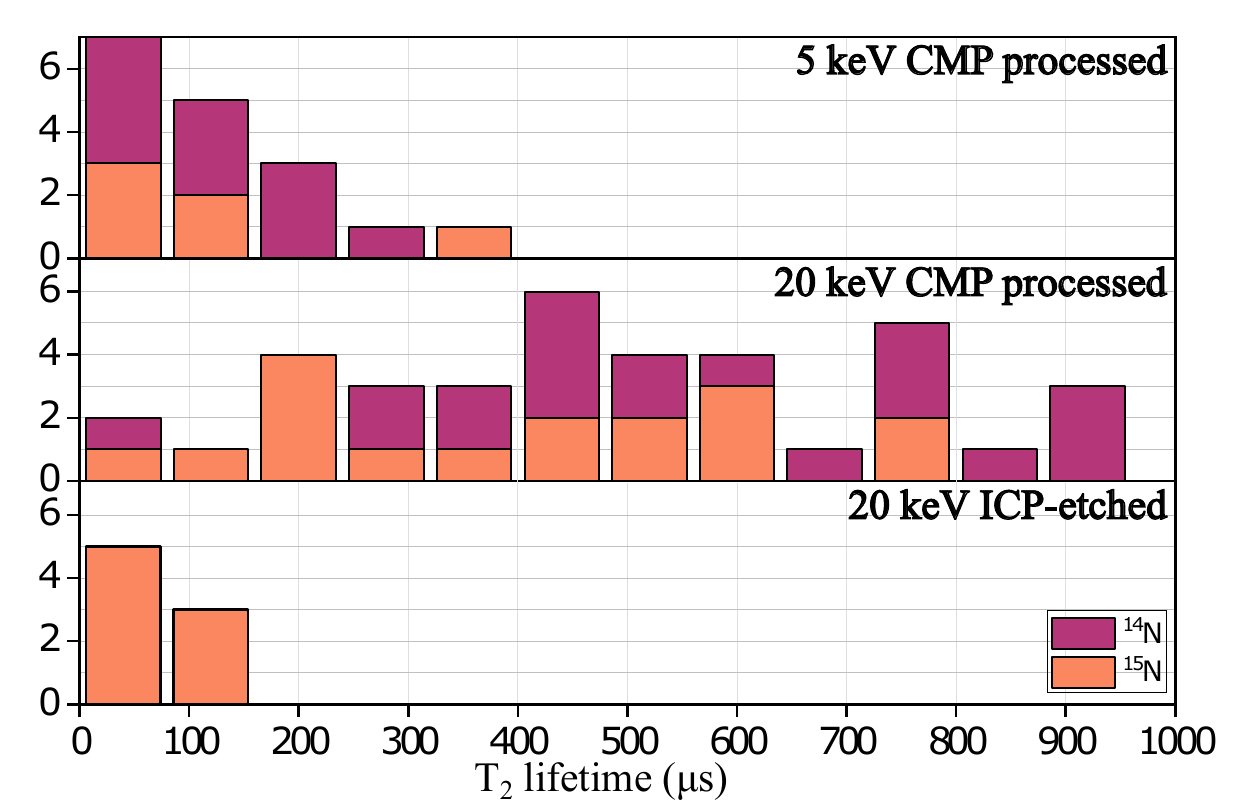}
    \caption{\label{fig:lifetime_histograms}The \ttwo{} times of all three samples, which includes lifetimes from both $^{14}$NV and $^{15}$NV. We observe \ce{13C}-limited spin lifetimes in the ICP+CMP-processed sample implanted at \SI{20}{keV}. The mean spin lifetimes in the ICP+CMP-processed sample are significantly longer than those we measured in a purely ICP-processed sample implanted and annealed under identical conditions.}
\end{figure}

The spin lifetimes show a clear trend with respect to implantation energy (and therefore depth) of the NV centre, with shallow NV centres demonstrating significantly shorter T$_2$. This suggests that surface effects are the primary source of decoherence, as previously observed. \cite{sangtawesin} At \SI{20}{keV}, a significant proportion possess spin lifetimes which are limited by the \ce{^{13}C} nuclear spin bath ($\gtrapprox\SI{700}{\micro\second}$
Fig~\ref{fig:lifetime_histograms}).\cite{stephen2019,stanwix2010,zhao2012} To our knowledge, these are the longest T$_2$ lifetimes reported for implanted NV centers, indicating that CMP is capable of producing surfaces with a subsurface damage level which is not limiting natural abundance spin lifetimes. 

In order to compare the spin lifetimes measured in the ICP+CMP-processed sample with the more conventional ICP-RIE surface processing, we also measured spin lifetimes for \SI{20}{keV}-implanted NV in the ICP-processed sample (which underwent no subsequent CMP processing). The measured T$_2$ times from this sample are significantly shorter than those from CMP treated samples, despite being implanted, annealed, and surface terminated under identical conditions. Considering only the implanted NVs (i.e. those containing $^{15}$N), the CMP treated sample shows an increase in the median of \SI{355}{\micro\second} and an increase in the mean of \SI{334}{\micro\second} compared to the ICP-RIE treated sample.

These results are in contrast to those published by Braunbeck et al, \cite{braunbeck} where no relation between spin lifetimes and surface polishing techniques was observed. However, we believe that not enough material was removed during the ICP-RIE and CMP treatments, leaving behind scaife-induced sub-surface damage that limited the T$_2$ times and prevented a true measurement of a CMP-processed surface.

\section{Conclusions}
Subsurface damage is a known problem in the processing of diamond for high-sensitivity applications such as quantum technologies and electronics. The present approach taken in academic and low-volume settings is to use ICP-RIE to process samples in small numbers. This requires high-capital equipment and as a result of the low etch rates typical in diamond for low roughness finishes, also results in low throughput.\cite{Toros2020} 

CMP offers a scalable route to volume production of smooth surfaces. In this work, we have exploited the sensitivity of the NV center to its environment to demonstrate that CMP processing also produces low subsurface-damage surfaces suitable for near-surface sensing and electronics applications. In our measurements, a significant proportion of implanted NV centers approach the limit imposed by the \ce{^{13}C} spin bath present in natural abundance samples. We measure median and mean spin lifetimes in ICP+CMP-processed surfaces which are significantly longer than those measured using a standard cyclical ICP-RIE-processed sample which was implanted, annealed, and terminated under identical conditions. 

Further work is required to ameliorate or prevent the sample rounding which is inherent to simple realizations of CMP on small, square samples. 

\begin{acknowledgments}
We acknowledge the support of ESPRC Grant No. EP/V056778/1.
\end{acknowledgments}

\clearpage

\appendix

\section{Sample Preparation}
\label{sec:sampleprep}
For this study, we employed five commercial electronic grade CVD grown diamonds provided by Element Six. Three samples (a,b,c) were treated with a pure \ce{O_2} ICP-RIE process to remove approximately \SI{10}{\micro\meter} of material. Each of these samples was then further processed with CMP (details of slurry and abrasives in main text) for between \num{50} and \SI{100}{\hour} to remove a further \SIrange{1}{10}{\micro\meter} of material. One sample (d) was processed using a cyclical \ce{O_2 /ArCl_2} ICP-RIE process to remove \SI{20}{\micro\meter} of material. The final sample (e) was used only for AFM measurements to provide a reference for the as-received surface roughness and surface texture.

\begin{table}[htbp]
\begin{tabular}{lp{2.5cm}p{3cm}}
Sample & \ce{^{15}N} energy (keV) & Notes\\
\hline
a & 20 & ICP+CMP processing\\
b & 10 & ICP+CMP processing\\
c & 5 & ICP+CMP processing\\
d & 20 & ICP processing\\
e & \multicolumn{1}{l}{N/A} & As-received for AFM
\end{tabular}
\end{table}

\begin{figure}[H]
\includegraphics[width=\columnwidth]{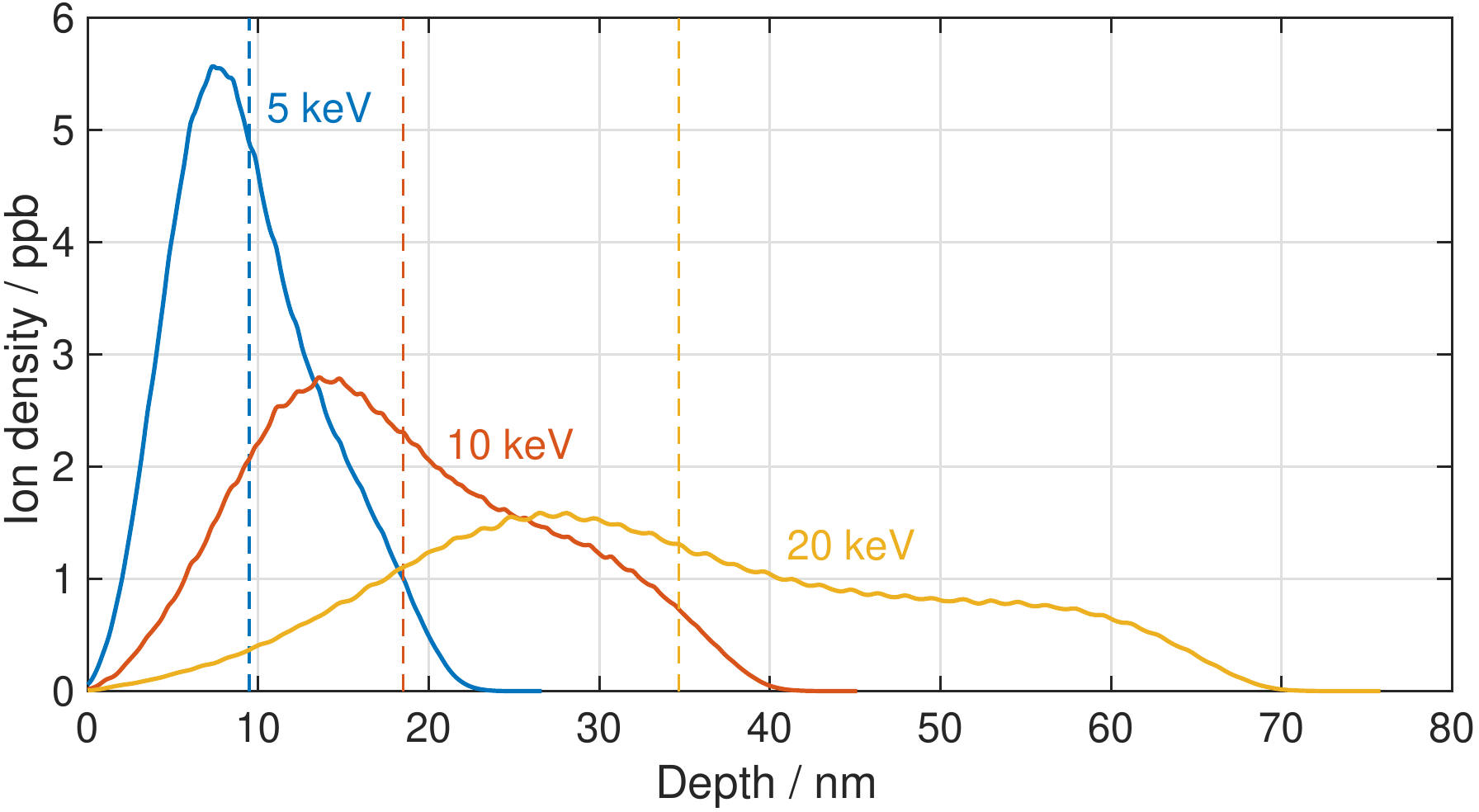}
\caption{\label{fig:trim} Final ion resting depths for \ce{^15N} ions implanted at the given energies, calculated by Crystal-TRIM. \cite{posselt1994}}
\end{figure}

Samples a \& d, b, and c, were implanted with \ce{^{15}N} ions at a sample tilt of \SI{7}{\degree} and energies of \num{20}, \num{10}, and \SI{5}{keV}, respectively. All samples were implanted to a total dose of  \SI{1e9}{\per\centi\meter\squared}. The mean depths of the implanted ions were calculated using Crystal-TRIM as \SI{34}{\nano\meter}, \SI{18}{\nano\meter}, and \SI{9}{\nano\meter} respectively (Fig.~\ref{fig:trim}). Following implantation, samples a---d were annealed in a nitrogen environment at \SI{400}{^\circ C} for 4 hours, \SI{800}{^\circ C} for 2 hours, and finally \SI{1200}{^\circ C} for 2 hours, based on the recipe given in \cite{Chu2014a}. 

Finally, all samples were cleaned with a Piranha solution (1:3 \ce{H2O2}:\ce{H2SO4}) and subsequently terminated under standard atmosphere with the sample held at \SI{465}{\celsius} for two hours in order to oxygen terminate.\cite{sangtawesin} This treatment was repeated when needed.


\section{Confocal Microscope and Pulsed Measurements}
\label{sec:confocal}
All spin lifetime measurements were taken using a home-built \SI{532}{\nano\meter} confocal microscope equipped with a Zeiss $100\times$, 1.4~NA magnification oil objective, and Immersol 518F immersion oil.

The initialization / readout excitation is supplied by a Laser Quantum Gem 532. The power and digital switching is controlled using an Isomet 1250C AOM, and the laser is circularly polarised to ensure equal excitation efficiency for all NV orientations. The light received from the diamond is filtered using a \SI{532}{nm} notch filter and a \SI{633}{nm} longpass filter. The filtered emission is subsequently split by a 50:50 beamsplitter and detected by two Excelitas SPCM-NIR single photon counting modules .

Microwaves were generated by a Keysight N5172B, amplified using a MiniCircuits ZHL-16W-43-S+ and delivered to the NVs using a \SI{20}{\micro\meter} diameter wire, approximately \SI{20}{\micro\meter} from the implanted NVs, terminated with \SI{50}{\ohm} terminator. The external magnetic field was applied using a permanent magnet.

\begin{figure}[hbtp]
\hspace*{-1.25cm}
\includegraphics[scale=0.4]{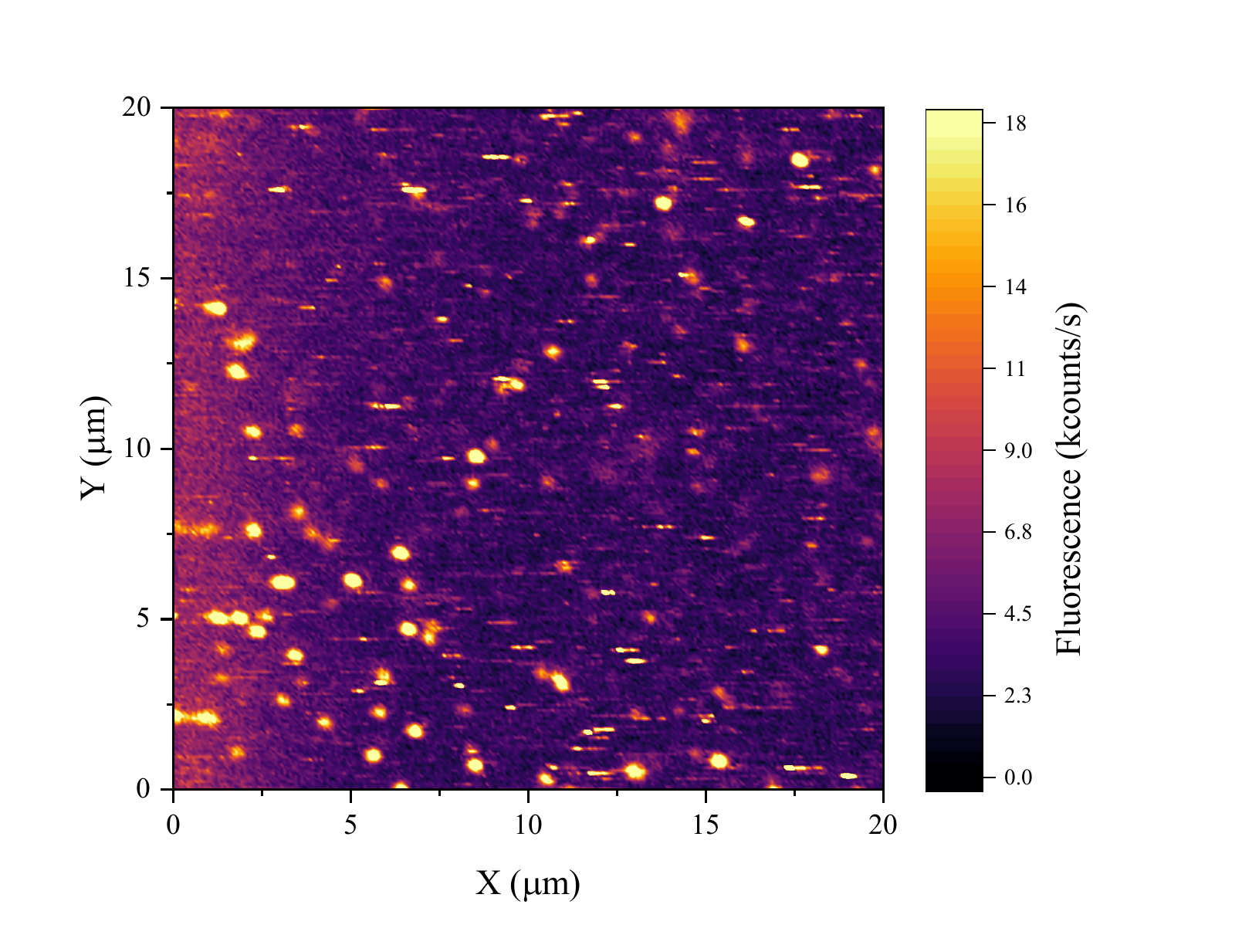}
\caption{\label{fig:confocalimage2} Confocal image of single NVs in a CMP treated sample.}
\end{figure}

Hanbury-Brown-Twiss (HBT) measurements were collected for all sampled NVs. Confirmation that each source is a single NV is given when the normalised HBT peak is $<0.5$.

\begin{figure}[hbtp]
\hspace*{-1cm}
\includegraphics[scale=0.4]{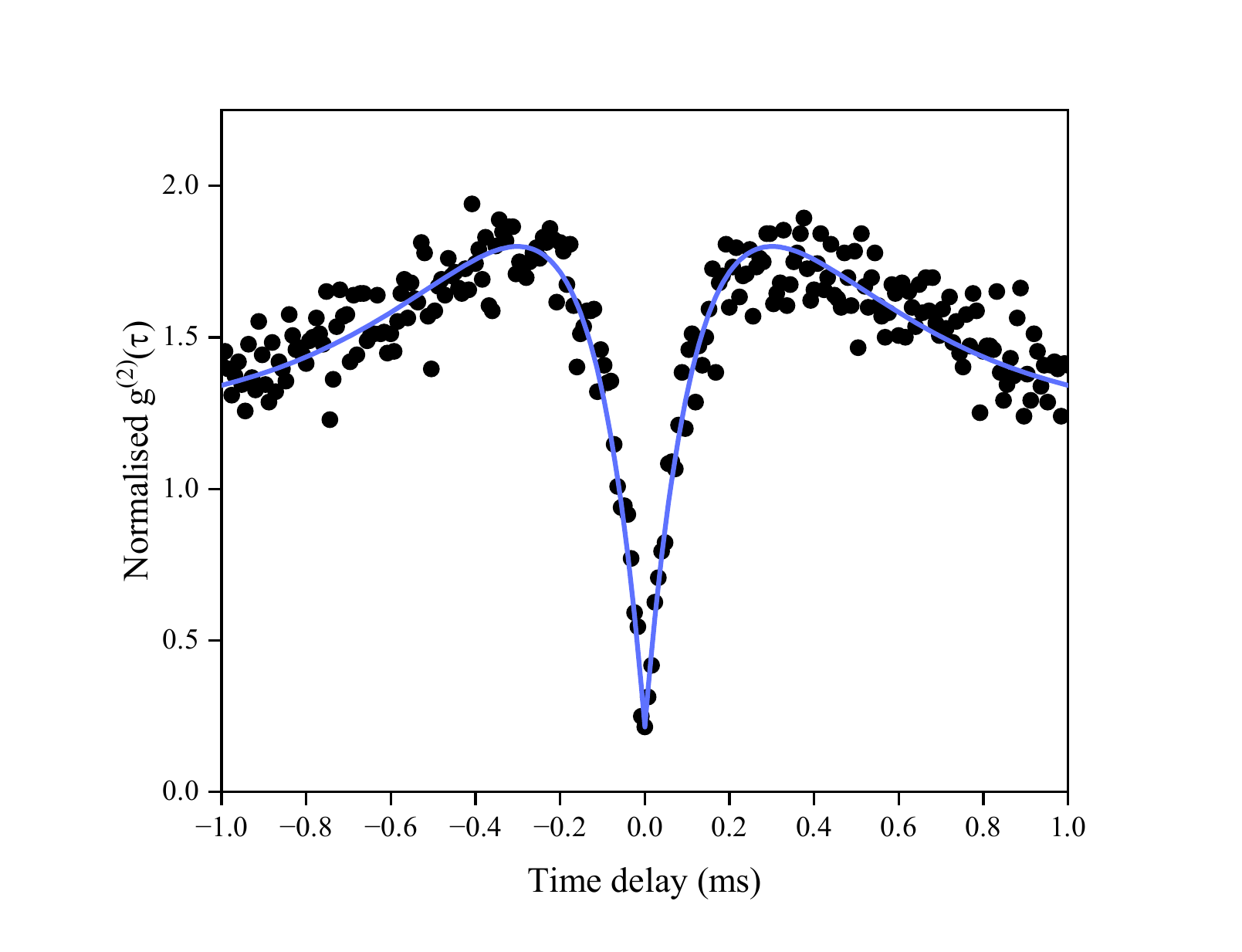}
\caption{\label{fig:confocalimage3} HBT measurement from a single NV, fitted with a second order autocorrelation function.}
\end{figure}

The nitrogen isotopes were determined using a Ramsey pulse sequence and the \ttwo{} times of each NV were measured using a Hahn echo pulse sequence, generated using a Swabian Instruments Pulse Streamer 8/2. A Ramsey sequence consists of a \SI{532}{\nano\meter} laser pulse to initialise the spin state into $\ket{m_s=0}$ followed by anmicrowave pulse of length $\frac{\pi}{2}$. Following a length of time of $\tau_{Ram}$, during which the spin can dephase, a second $\frac{\pi}{2}$ microwave pulse is applied before a final \SI{532}{nm} laser pulse to readout the spin state. All Ramsey microwave pulses were detuned by \SI{5}{\mega\hertz} to improve isotope identification via Fourier transform.  During a Hahn echo sequence, the spin is initialised in the same way (using an on-resonance pulse) but after $\tau_{Hahn}$ of dephasing time an additional $\pi$ length microwave pulse is applied, and a spin echo occurs $\tau_{Hahn}$ after the $\pi$ pulse, and the spin state is readout with a final \SI{532}{nm} laser pulse. 

\nocite{*}
\bibliography{aipsamp}

\end{document}